\documentclass{phbauth}
\usepackage{graphicx}
\newcommand{\text}[1]{{\mathrm{#1}}}
\renewcommand{\vec}{\mathbf}

\begin{document}
\begin{frontmatter}
\title{Quantum Hall droplets in coupled lateral quantum dots}

\author{Ramin M. Abolfath,}
\author{Pawel Hawrylak,}
\author{Michel Pioro-Ladriere,}
\author{and Andy Sachrajda}

\address{
Institute for Microstructural Sciences,
National Research Council of Canada,
Ottawa, K1A 0R6, Canada}


\begin{abstract}
We present a theory and Coulomb and Spin Blockade spectroscopy 
experiments on quantum Hall droplets with controlled
electron numbers (N1,N2) in laterally coupled gated quantum dots. 
The theory is based on the configuration interaction method (CI) coupled with 
the unrestricted Hartree-Fock (URHF) basis. It allows us to calculate the magnetic  
field evolution of ground and excited states of coupled quantum dots 
with large electron numbers. The method is applied to the spin transitions
in the (5,5) droplet. Preliminary experimental results demonstrate the
creation of the (5,5) droplet and its Spin Blockade spectra.
\end{abstract}
\end{frontmatter}



\begin{figure}
\begin{center}
\vspace{1cm}
\includegraphics[width=0.98\linewidth]{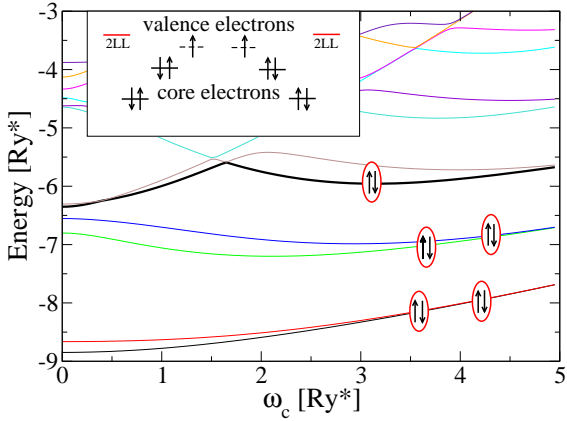}
\caption{Double dot single particle spectrum vs. cyclotron energy.
}
\label{E_sp}
\end{center}
\end{figure}

The coupled lateral quantum dots form artificial molecules 
with each dot playing the role of an artificial 
atom\cite{palacios,kouwenhoven,MichelPRL93,petta}.
In strong magnetic field electrons form quantum Hall droplet in each
quantum dot. In a double dot one can couple the quantum Hall
droplets in a controlled way, and at filling factor $\nu=2$  
effectively reduce the many-electron-double dot system to a 
two-level molecule\cite{MichelPRL93}, as illustrated in the inset
to Fig.\ref{E_sp}.  The singlet-triplet spin transitions
of a two-level molecule have potential application 
as quantum gates. \cite{brum,loss-divincenzo,burkard-loss,HuDasSarma} 
Here we present a theory of spin transitions in many-electron double quantum dots 
in a magnetic field
and preliminary  results of Spin Blockade spectroscopy (SB)
on double dots with controlled electron numbers (N1,N2)
in dot 1 and dot 2.

Our theory is based on effective mass envelop function to describe the
confined electrons in quantum dots. We consider electron motion to be
quasi-two-dimensional and coupled to the perpendicular external magnetic 
field  by vector potential $A$.  With the total number of electrons $N=N1+N2$
the quantum dot molecule Hamiltonian can be written as:

\begin{equation}
H = \sum_{i=1}^N  T_i
+ \frac{e^2}{2\epsilon}\sum_{i \neq j}\frac{1}{|\vec{r}_i - \vec{r}_j|},
\end{equation}

where $T=\frac{1}{2m^*}\left(\frac{\hbar}{i}\nabla 
+ \frac{e}{c} A(\vec{r})\right)^2 + V(x,y)$ is the 
one electron Hamiltonian with $V(\vec{r})$  the 
quantum dot molecule confining potential, 
$m^*$  the conduction-electron effective mass, and
$\epsilon$ the host semiconductor dielectric constant. 
The Zeeman spin splitting (very small for GaAs) is neglected here.
In what follows we use GaAs effective atomic energy and length units with  $Ry^*=5.93 meV$, and 
$a^*_0 = 9.79 nm$.

The double quantum dot potential $V(x,y)$ defined by electrostatic gates is characterized by
two potential minima. With our focus on electronic correlations, 
we parameterize electrostatic potential of a general class of coupled
 quantum dots by a sum of three Guassians  \cite{HuDasSarma} 
$V(x,y)=V_1~ \exp[{-\frac{(x+a)^2+y^2}{\Delta^2}}]
        +V_2~ \exp[{-\frac{(x-a)^2+y^2}{\Delta^2}}]
+V_p \exp[{-\frac{x^2}{\Delta_{Px}^2}-\frac{y^2}{\Delta_{Py}^2}}]$.
Here $V_1,V_2$ describe the depth of the left and righ quantum dot minima
located at $x=-a,y=0$ and $x=+a,y=0$, and $V_p$ is the plunger gate potential
controlled by the central gate. For identical dots, $V_1=V_2$ , and confining
potential  exhibits inversion symmetry. In what follows we will 
parameterize it by $V_0=-10, a=2, \Delta=2.5$, 
and $\Delta_{Px}=0.3$, 
$\Delta_{Py}=2.5$, in effective atomic units. 
$V_p$, which controls the potential barrier ,
is varied between zero and $10 Ry^*$, independent 
of the locations of the quantum dots.

The potential of each isolated dot is a single Gaussian potential. Expanding
it in the vicinity of the minimum yields a
parabolic potential $V(r)=1/(2 m^*) \omega_0 r^2$ with the strength 
$\omega_0=2\sqrt{|V_0|/\Delta^2}$.  
The low energy spectrum of each dot corresponds to two
harmonic oscillators with eigen-energies
$\varepsilon_{nm}=\hbar\omega_+(n+1/2)+\hbar\omega_-(m+1/2)$.
Here  $\omega_\pm = \sqrt{\omega_0^2 + \omega_c^2/4} \pm \omega_c/2$, $\omega_c$
is the cyclotron energy, and
$n,m=0,1,2,...$.
With increasing magnetic field the $\hbar\omega_-$ decreases to zero
while $\hbar\omega_+$ approaches the cyclotron energy $\omega_c$, and
the states $|m,n\rangle$ evolve into the nth Landau level.
The $\nu=2$ spin singlet quantum Hall droplet is formed 
if 2N electrons occupy the 
N successive $ | m,n=0 \rangle$ lowest Landau level (LLL) orbitals. 
When  extra  $2N+1$th electron
is added it  occupies the edge orbital $|m=N,n=0 \rangle$. 
These $(2\times 2+1,2 \times 2+1)$ configurations, for the
two isolated dots, are shown as inset in Fig.\ref{E_sp}. 
In each isolated dot increasing the
magnetic field leads to spin flips while 
decreasing the magnetic field lowers the energy of the edge 
$|m=N,n=0 \rangle$ orbital
with respect to the lowest unoccupied center orbital $|m=0,n=1 \rangle$ of the 
second Landau level (2LL). At a critical magnetic field a LL crossing occurs 
and the electron transfers from the edge orbital to central orbital,
leading to redistribution of electrons from the edge to center.\cite{WKH}

We now turn to the description of coupled dots in strong magnetic field
($n=0$).
For weak coupling we expect the $|m; 1 \rangle$ orbitals of the first dot
to be coupled with the $|m; 2 \rangle$ orbitals of the second dot,
and form the symmetric and anti-symmetric orbitals 
$|m; \pm \rangle = |m; 1 \rangle \pm |m; 2 \rangle$.
Therefore in high magnetic field we expect the formation of
shells of closely spaced pairs of levels. This is illustrated
in  Fig. \ref{E_sp} which shows the magnetic field evolution of 
the numerically calculated single particle spectrum of a  double dot.
The spectrum is calculated accurately by discretizing in real space 
the single particle Hamiltonian $T$ 
and applying special gauge transformation. The resulting large
matrices are diagonalized  using conjugate gradient algorithms. 

At zero magnetic field Fig. \ref{E_sp} shows the formation of 
 hybridized S, P, and D shells.
In high magnetic field the pairs of closely spaced
levels $ |m,\pm \rangle $ separated by $\approx \omega_+$ are clearly visible. 

We can now populate the $ |m,\pm \rangle $ electronic shells and form
states equivalent to   $\nu=2$ droplets in coupled quantum dots.
The half-filled shells correspond to $(N1,N2)=(1,1),(3,3),(5,5),(7,7)..$ 
while filled shells correspond to $(N1,N2)=(2,2),(4,4),(6,6),(8,8)..$ 
configurations.
The population of the (5,5) configuration is shown in Fig.\ref{E_sp}.
We expect the half and fully filled shells to have special electronic properties.
In particular, the half filled shells offer the possibility of singlet-triplet
transitions. For the shells (5,5) and up  we expect to be able
to move  the valence electrons from the edge orbitals to the center orbitals.
The crossing of the edge and center orbitals is visible as a cusp in the energy of the
fifth molecular orbital shown as a bold line in Fig.\ref{E_sp}.

\begin{figure}
\begin{center}
\includegraphics[width=0.98\linewidth]{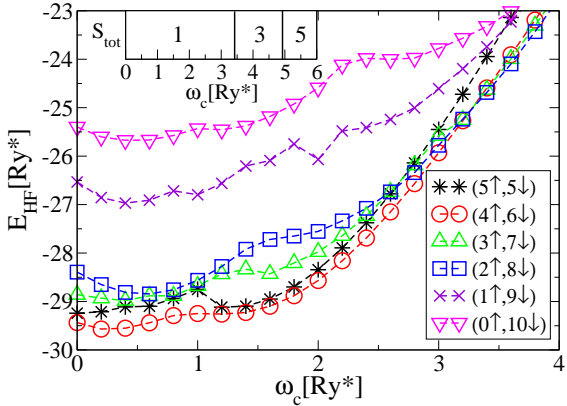}
\noindent
\caption{
URHFA ground state energy vs. cyclotron energy is shown
for a quantum dot molecule.
Spin transitions from $S=1$ to $S=3$, and from $S=3$ to $S=5$
occur at $\hbar\omega_c=3.4 Ry^*$, and $\hbar\omega_c=4.9 Ry^*$ (inset).
}
\label{URHF_10e}
\end{center}
\end{figure}

We now proceed to include electron-electron interactions in two steps:
direct and exchange interaction using unrestricted Hartree-Fock approximation
(URHF), and correlations using URHF basis in the configuration interaction method (URHF-CI). 
The spin-dependent HF orbitals $|\varphi_{i\sigma}\rangle $ are obtained 
from  the $N_l$ non-interacting
single particle orbitals $|\tilde{\varphi}_{\alpha}\rangle$,
 energy spectrum of which
is shown in Fig.\ref{E_sp},
 by the transformation
$|\varphi_{i\sigma}\rangle = \sum_{\alpha=1}^{N_l} a^{(i)}_{\alpha\sigma} 
|\tilde{\varphi}_{\alpha}\rangle$.
The variational parameters $a^{(i)}_{\alpha\sigma}$
are solutions of self-consistent Pople-Nesbet 
equations \cite{Szabo_book}:

\begin{eqnarray}
&&\sum_{\gamma=1}^{N_l}
\{\tilde{\epsilon}_\mu\delta_{\gamma\mu}+
\sum_{\alpha,\beta=1}^{N_l}\tilde{V}_{\mu\alpha\beta\gamma}
[\sum_{j=1}^{N_\uparrow}
a^{*(j)}_{\alpha\uparrow} a_{\beta\uparrow}^{(j)} 
+ \sum_{j=1}^{N_\downarrow} \nonumber \\ &&
a^{*(j)}_{\alpha\downarrow} a_{\beta\downarrow}^{(j)} ] 
-\tilde{V}_{\mu\alpha\gamma\beta}
\sum_{j=1}^{N_\uparrow}
a^{*(j)}_{\alpha\uparrow} a_{\beta\uparrow}^{(j)}  
\} a_{\gamma\uparrow}^{(i)} 
= \epsilon_{i\uparrow} ~ a_{\mu\uparrow}^{(i)},
\label{urhfeq1}
\end{eqnarray}

where $\tilde{V}_{\alpha\beta\mu\nu}$ are Coulomb matrix elements
calculated using non-interacting single particle states.
A similar equation holds for spin down electrons.
The calculations are carried out for
all possible total spin $S_z$ 
configurations.
The calculated total energies for the $N=10$ electrons with $S_z=0,1,..,5$  in a
magnetic field are shown in Fig. \ref{URHF_10e}.
We find the lowest energy state to correspond to $S_z=1$  up to $\hbar\omega_c=3.4Ry^*$,
 $S_z=3$ for $  3.4Ry^* < \hbar\omega_c < 4.9Ry^*$, and $S_z=5$ for $\hbar\omega_c > 4.9Ry^*$ .
The predicted by URHFA total spin evolution with magnetic field  is shown 
as inset in Fig. \ref{URHF_10e}. It is remarkable to note that the $S=0$ singlet (5,5)
state is never a ground state, and the URHF picture is drastically different from the
noninteraction picture presented in Fig.\ref{E_sp}. The situation is improved by
including the electron correlation energy. 


\begin{figure}
\begin{center}\vspace{-1cm}
\includegraphics[width=0.98\linewidth]{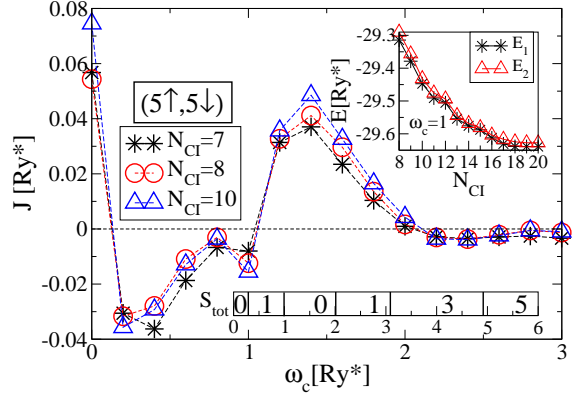}
\noindent
\caption{
The energy difference J between the triplet and the singlet 
ground states of the (5,5) quantum Hall droplets from URHF-CI
as a function of cyclotron energy $\omega_c$.
The total spin $S_{\rm tot}$ in magnetic field is shown in the lower inset.
The ground state and first excited state energy as a function of $N_{S}$,
the number of HF basis used in CI calculation, is shown in the upper inset.
}
\label{J4}
\end{center}\vspace{0.5cm}
\end{figure}

Correlations are included via CI method. Denoting the creation (annihilation) 
operators for URHF quasi-particles by $c^\dagger_{i}$ ($c_{i}$) with
the index $i$ representing the combined spin-orbit quantum numbers,
the many body Hamiltonian of the interacting system 
can be written as:
\begin{equation}
H=\sum_{ij} 
\langle i | T | j \rangle 
c^\dagger_{i} c_{j} +
\frac{1}{2}\sum_{ijkl}
V_{ijkl}
c^\dagger_{i} c^\dagger_{j} 
c_{k} c_{l} ,
\label{multiparticle}
\end{equation}       
where 
$
\langle i | T | j \rangle = \epsilon_i \delta_{ij} -
\langle i | V_H + V_X | j \rangle,
$
$V_{ijkl}$ are the Coulomb matrix elements in the URHF basis,
$\epsilon_i$ are the URHF eigenenegies, $V_H$ and $V_X$ are the 
Hartree and exchange operators.
The Hamiltonian matrix is constructed in the basis of configurations, and
diagonalized using conjugated gradient methods for different total $S_z$.
The convergence of CI calculation for the (5,5)
droplets has been checked by increasing
the URHF basis up to $N_{S}=20$,
associated with $240~374~016$ configurations, with results
shown as inset in Fig. \ref{J4}.

\begin{figure}
\begin{center}
\includegraphics[width=0.98\linewidth]{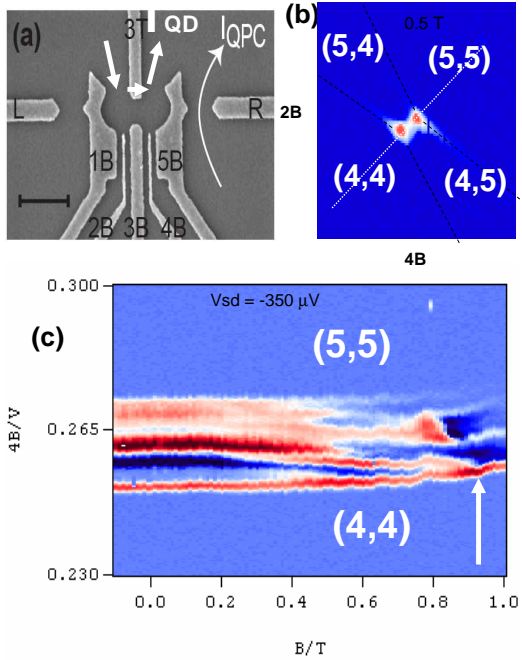}
\noindent
\caption{(a) double dot gate layout, (b) transport through the ddot around 
(5,5) configuration at B=0.5T, and (c) current stripe at finite source-drain voltage
as a function of magnetic field.
}
\label{fig4exp}
\end{center}
\end{figure}

The results of calculated exchange interaction 
$J\equiv E_{\rm triplet} - E_{\rm singlet}$, 
and the spin phase diagram of the $(5,5)$-droplet (inset)
in magnetic field are shown in Fig. \ref{J4}. 
Unlike in URHF, for magnetic fields less than 
$3.2 Ry^*$, the ground state of the $(5,5)$-droplet oscillates between
$S=0$ singlet state and $S=1$ triplet state. Correlations
restored the 
$\nu=2$ quantum Hall droplet spin singlet ($S=0$) phase
as the phase within $1<\hbar\omega_c<2$.
This phase is unstable against spin triplet state at $\hbar\omega_c = 2$
and at $\hbar\omega_c \approx 1$. The spin singlet-triplet transition at
$\hbar\omega_c = 2$ is equivalent to the magnetic field induced
singlet-triplet transition in a two-electron double dot\cite{burkard-loss}.
The spin singlet-triplet transition at $\hbar\omega_c \approx 1$ is associated
with degeneracy of  the crossing LLL edge  and 2LL center orbitals.

At $\hbar\omega_c=3.2 Ry^*$ spin state with $S=3$
is formed and at $\hbar\omega_c=4.8 Ry^*$ a fully polarized state 
developes. The full spin evolution is summarized in the inset in Fig. \ref{J4},
and will be discussed in detail in the future.

In order to test theoretical results we have designed a double dot
with tunable electron numbers. The gate layout is shown in Fig.\ref{fig4exp}a.
Electrons are counted using both Coulomb and Spin Blockade spectroscopy
and  charge detection. Fig.\ref{fig4exp}b shows current at B=0.5T 
as a function of the 2B and 4B gates for device
with (5,5),(4,5),(5,4) and (4,4) electron configurations.
In Fig.\ref{fig4exp}c we show the current stripe as a function of the magnetic field
for a fixed source-drain voltage. The structures in the stripe correspond to
excited states between  the (4,4) and (5,5) configurations. The point 
indicated by the arrow suggests a transition in the double dot. We speculate that this
transition corresponds to moving the 9th electron from the edge to the center 
orbital as shown in Fig.\ref{E_sp}. Much more work is needed to identify spin 
transitions in the (5,5) droplet. 

Acknowledgement. R.A. and P.H. acknowledge support by the NRC High 
Performance Computing project.



\end{document}